\documentclass{article}
\newcommand{\bfr}{\begin{flushright}}
\newcommand{\efr}{\end{flushright}}
 
\begin{document}
\title{Zero Modes in Vortex-Fermion System with Compact Extra Space
}
\author{ 
Atsushi Nakamula\\
Department of Physics, Tokyo Metropolitan University,\\
Setagaya-ku, Tokyo 158, Japan\\
and\\
Kiyoshi Shiraishi\\
Institute for Nuclear Study, University of Tokyo, \\
Midori-cho, Tanashi,
Tokyo 188, Japan
}
\date{Modern Physics Letters {\bf A5}, No.~14 (1990) pp.~1109--1117
}
\maketitle
\begin{abstract}
The existence of fermionic zero modes is shown in the presence of
vortex configuration of pure $SU(2)$ gauge field on the manifold $M_4 \times S^2$.
From the perspective of four-dimensional effective theory, these zero
modes are almost the same as the Jackiw-Rossi type zero modes of the
vortex-fermion system.
\end{abstract}

\section{Introduction}
To construct a unified theory of fundamental interactions including gravity,
there exists a promising possibility in higher-dimensional theories, namely, string
theories~\cite{1} which are formulated naturally in space-time dimensions greater than
four. Although in such theories we must solve problems related to the dynamics of
compactification of extra dimensions, we can immediately consider low energy
(i.e., below the Planck scale) effective field theories in four space-time after
compactiffcation.

From the perspective of these effective theories, it is permissible that vector
fields have non-vanishing vacuum expectation values (VEV) for their extra dimensional
components since they do not break four-dimensional Poincar\'e invariance of the theory. Under
these circumstances, the VEV of the gauge field on extra-space manifold are quite important.
For example, when the gauge field configuration has non-zero field strength, they can be a
source to stabilize the extra space compact; otherwise fermions generally must gain masses of
order Planck scale in four dimensions. The finite field strength may be a candidate for
cancellation of such large masses. Furthermore, in the case that the gauge group under
consideration is non-Abelian, the existence of the non-vanishing components can be a seed for
gauge symmetry breaking. A profitable fact in this mechanism to break large gauge
symmetry is that it is not necessary to introduce an extra scalar (Higgs) field with an
appropriate shape of its potential. In other words, in such mechanism, the role of the
Higgs field which gives the gauge field mass is imposed on the gauge field itself.

With this type of symmetry breaking mechanism we may consider associated
phase transitions and a simultaneous production of topological defects. For a
concrete example of such defects, Lee, Holman and Kolb \cite{2} constructed a domain
wall solution of pure $SU(3)$ gauge theory in seven dimensions. And the present
authors have recently found a vortex-type solution \cite{3} to the pure Yang-Mills ($SU(2)$)
equation of motion which couples with gravity on $M_4\times S^2$, where $M_4$ is a large four
space-time and $S^2$ is an extra two-sphere. From the four-dimensional point of view,
this configuration is almost an isolated infinitely long cosmic string of an Abelian-Higgs
theory.\cite{4} The structure of this vortex solution has quite a curious feature that in the
core of the vortex the $SU(2)$ gauge symmetry is breaking to $U(1)$, while outside the core the
symmetry is restored,\cite{5} namely, the symmetry structure is converse to the Abelian-Higgs
vortex.\cite{6} In both examples of defects, the stability of field configurations is
guaranteed by non-trivial topological winding number, i.e.,
$\sim\int{\rm tr }F\wedge F$.

Now, when we introduce fermions into the system with topological defects, some
curious properties appear. In particular, the existence of fermionic zero modes are
well investigated in various systems which possess topological objects as backgrounds, e.g.,
magnetic monopole,\cite{7} vortex,\cite{8} instanton,\cite{9} etc. In this letter, we consider
fermion zero modes against the background of the $SU(2)$ vortex solution in six dimensions. In
the four-dimensional cosmic string model, the existence of such zero modes is crucial for a
mechanism in which the string behaves as superconducting wire.\cite{10} Generally, to
investigate the fermion zero modes we have two directions: (i) From an index theorem we can get
the difference in number between the left- and right-handed zero modes of an associated Dirac
operator. (ii) By solving the Dirac equation directly, we can explicitly show the number of
zero modes and their chiralities. This calculation is done in the case of an Abelian-Higgs
vortex in $2+1$ dimensions by Jackiw and Rossi.\cite{8} We will consider these two methods.

At the beginning, we give the background $SU(2)$ gauge field of vortex type in six
dimensions.\cite{3} As mentioned above, the space-time topology is taken to be $M_4\times S^2$.
Throughout present consideration, $M4$ is set to be flat and the geometry is fixed, i.e.,
gravity is neglected, although in the original model,\cite{3} the dynamics of gravity was
included. Since the vortex system has an axial symmetry we use a cylindrical
coordinate of $M_4$:
\begin{equation}
ds^2=-dt^2+dz^2+dr^2+r^2d\psi^2+b^2(d\theta^2+\sin^2\theta d\phi^2) \,,
\label{1}
\end{equation}
where $b$ is a constant radius of the extra two-sphere. The vortex configuration of
pure $SU(2)$ gauge field is the following solution to the Yang-Mills equation. At the
core of the string, the configuration should have a finite energy density, so we place
a monopole configuration \cite{11} on extra space there, namely,
\begin{eqnarray}
& &{\bf A}_\theta^M=0\,,\nonumber \\
& &{\bf A}_\phi^M=(1-\cos\theta)\frac{1}{2}\left(
\begin{array}{rr}
1&0\\
0&-1
\end{array}
\right)\,.
\label{2}
\end{eqnarray}
This can be smoothly connected with a trivial vacuum. We write the configuration
via real functions $\Phi_1$ and $\Phi_2$ which are independent of the internal coordinates:
\begin{eqnarray}
{\bf A}_\theta&=&\Phi_1\frac{1}{2}\left(
\begin{array}{cc}
0&-ie^{-i\phi}\\
ie^{i\phi}&0
\end{array}
\right)+\Phi_2\frac{1}{2}\left(
\begin{array}{cc}
0&e^{-i\phi}\\
e^{i\phi}&0
\end{array}
\right)\,,\nonumber \\
{\bf A}_\phi&=&-\Phi_1\frac{1}{2}\left(
\begin{array}{cc}
0&e^{-i\phi}\\
e^{i\phi}&0
\end{array}
\right)\sin\theta+\Phi_2\frac{1}{2}\left(
\begin{array}{cc}
0&-ie^{-i\phi}\\
ie^{i\phi}&0
\end{array}
\right)\sin\theta\nonumber \\
& &+(1-\cos\theta)\frac{1}{2}\left(
\begin{array}{rr}
1&0\\
0&-1
\end{array}
\right)\,,
\label{3}
\end{eqnarray}
where the case $\Phi_1=\Phi_2=0$ is trivially the monopole, whereas the case
$|\Phi_1+i\Phi_2|=1$ can be shown to be gauge equivalent to the trivial vacuum. These extra
space components can be represented as a complex scalar field $\Phi\equiv\Phi_1+i\Phi_2$.
Further, we place an ansatz on the components,
\begin{equation}
{\bf A}_\mu=\frac{1}{2}A_\mu\left(
\begin{array}{rr}
1&0\\
0&-1
\end{array}
\right)\,,
\label{4}
\end{equation}
where Greek indices run over $M_4$ coordinates. These are commutable with Eq.~(\ref{2})
and with each other. For the fields $\Phi$ and $A_\mu$, the Yang-Mills equation can be
reduced to that of an effective $U(1)$-scalar system with a ``Mexican hat'' potential.\cite{3}
Now we construct a vortex-type solution in large four space-time from this background gauge
field. By imposing an axial symmetry in four dimensions, we need only the azimuthal component
of the gauge field. So, for an $n$-vorticity case we take an ansatz on the background as
\begin{equation}
{\bf A}_\psi=n(P(r)-1)\frac{1}{2}\left(
\begin{array}{rr}
1&0\\
0&-1
\end{array}
\right)\,,\quad
\Phi=X(r)\,e^{in\psi}\, .
\label{5}
\end{equation}
Boundary conditions for the functions $P(r)$ and $X(r)$ are $P(0)=1$, $P(\infty)=0$, $X(0)=0$
and $X(\infty)=1$. In this situation, we found numerical solutions for $X(r)$ and $P(r)$ with
unit vorticity. The multivorticity case will be solved in the same way. The behavior
of this vortex solution is almost the same as that of the Abelian-Higgs vortex as
expected from the equation of motion. For details see Ref.~\cite{3}.

Now, we introduce a massless fermion in a fundamental representation of $SU(2)$
into this system. The Lagrangian is written as
\begin{equation}
{\cal L}=\sqrt{-g}\left(\frac{1}{4e^2}{\rm tr
}F_{MN}F^{MN}+i\bar{\Psi}^aD\!\!\!\!/ \,\Psi^a\right)\,,
\label{6}
\end{equation}
where $M, N$ run over the six-dimensional coordinate and the $a$'s are summed over
the dimension of the fundamental representation. The field strength and a covariant derivative
for fundamental representation are defined as follows,
\begin{equation}
F_{MN}=\partial_M {\bf A}_N-\partial_N {\bf A}_M+i[{\bf A}_M, {\bf A}_N]\,,
\label{7}
\end{equation}
\begin{equation}
D_M=\nabla_M+i{\bf A}_M\, ,
\label{8}
\end{equation}
where $\nabla_M$ is a covariant derivative with respect to a gravitational connection.

First we consider the calculation of the index. As mentioned above, we can gain
some information of fermionic zero modes by an index theorem.

Let us remember the ordinary Abelian-Higgs vortex model as a similar case. In
the model of fermion-vortex system in $2+1$ dimensions,\cite{8} one of the crucial
conditions that the fermion has zero-energy modes is the existence of fermion-scalar (Yukawa)
couplings. From these, the fermion gains a mass in a true vacuum sector, i.e., far from the
defect, and in the neighborhood of the vortex there exist zero-energy modes which are
eigenstates of ``particle conjugation''. In this Abelian- Higgs model, the index theorem is
conjectured in Ref.~\cite{8} and verified by Weinberg,\cite{12} that is
\begin{equation}
{\cal G}=\frac{e}{\pi}\int d^2x\,F_{12}+\frac{1}{2\pi}\int_C
dl_i\frac{\varepsilon_{ab}\phi_a(D_i\phi_b)}{|\phi|^2}=
\frac{1}{2\pi}\int_C
dl_i\frac{\varepsilon_{ab}\phi_a(\partial_i\phi_b)}{|\phi|^2}\,,
\label{9}
\end{equation}
where we have used the fact $\int d^2x F_{12}=\int_C dl^i A_i$, $A_i$ is the $U(1)$ gauge
field, and $F_{ij}$ is the field strength. The left-hand side is an analytic index of an
associated differential operator, or, a generalized Dirac operator. The physical meaning of
this quantity is the difference in number between $+$ and $-$ eigenstates of particle
conjugation. This index theorem yields ${\cal G}=n$ in the $n$ vortex sector, so we find that
there are at least $|n|$ ``zero energy'' modes in the two-space on which the vortex lies.

On the other hand, in even dimensions of Euclidean signature there exists
another index theorem \cite{13} which says that an analytic index of the Dirac operator is
connected with a topological index of an associated manifold, ${\cal M}$, that is,
\begin{equation}
n_L-n_R=\frac{1}{32\pi^2}\int_{\cal M}{\rm tr } F\wedge F+
\frac{1}{96\pi^2}\int_{\cal M}{\rm tr } R\wedge R\,,
\label{10}
\end{equation}
where the left-hand side is the difference between the numbers of left- and right-
handed zero modes, which is the index of a spin complex. In the present pure $SU(2)$
vortex case, the fermion has only minimal gauge coupling which includes both
``gauge'' and ``scalar'' fields in the four space-time perspective. If we want to extract
some information in the present case from the topological index, we need only to
consider the index theorem (\ref{10}).

Now, we calculate the topological index of the vortex gauge field of the present
model. The vortex is an object which lies on the $(r, \psi)$ plane and on $S^2$ in this model,
so we are interested in the index theorem only for transverse directions including
an extra space, namely, the space spanned by coordinates $r, \psi, \theta$, and $\phi$. Then, we
divide the Dirac operator into two parts,
\begin{equation}
D\!\!\!\!/\,\Psi=(\Gamma^0\partial_t+\Gamma^3\partial_z)\Psi+D\!\!\!\!/\,{}_T\Psi\,,
\label{11}
\end{equation}
where $D\!\!\!\!/\,{}_T$ is the transverse Dirac operator. The transverse four-space is
partially compactified without boundary, namely, ${\bf R}^2\times S^2$ of constant curvature, so
$tr R ^ R - 0$. Incidentally, this holds in the case with dynamics of gravity. There-
fore the topological index comes from only a contribution of the $SU(2)$ gauge field
part. We obtain the index from Eq.~(\ref{10}):
\begin{equation}
\frac{1}{32\pi^2}\int_{\cal M}{\rm tr } F\wedge F=\int_0^\infty dr
\frac{d}{dr}(nP(1-X^2))=-n\,.
\label{12}
\end{equation}

The conclusion is
\begin{equation}
n_L-n_R=-n\,,
\label{13}
\end{equation}
where $n$ is the vorticity which appears in Eq.~(\ref{5}). So we find that there are at
least $|n|$ transverse zero modes in the n vortex sector. This result is almost the same
as the $2+1$ dimensional Abelian-Higgs vortex system. As mentioned above, in such
a system the index (\ref{9}) is the difference in number between ``particle conjugation''
eigenstates. However, in the present model the index we have obtained is the
quantity concerned with the chirality of the spinor in transverse space, ${\bf R}^2\times S^2$.
Of course these two quantities are closely related to each other.

So far we confirmed the existence of the transverse zero modes. From now on we
explicitly solve the Dirac equation to find the true number of zero modes and their
chiralities in the transverse space. The six-dimensional Dirac matrices are chosen
to be
\begin{eqnarray}
\Gamma^\mu&=&\gamma^\mu\otimes\sigma^1\,,\quad(\mu=0, 1, 2, 3)\nonumber \\
\Gamma^4&=&\gamma_5\otimes\sigma^1\,,\nonumber \\
\Gamma^5&=&1_4\otimes\sigma^2\,,
\label{14}
\end{eqnarray}
where $\gamma^\mu$'s, $\gamma_5$ and $1_4$ are four-dimensional Dirac, chiral and unit matrices
respectively, and $\sigma^i$'s are Pauli matrices. In this representation, a six-dimensional
chiral matrix becomes diagonal as
\begin{equation}
\Gamma_{\#}=1_4\otimes\sigma^3\, .
\label{15}
\end{equation}
The transverse Dirac operator can be written in terms of vielbein $e^m_\alpha$ and spin
connection $\omega_{m\beta\gamma}$ as
\begin{eqnarray}
D\!\!\!\!/\,{}_T&=&\Gamma^1\left\{\cos\psi\partial_r-\frac{\sin\psi}{r}(\partial_\psi+i{\bf
A}_\psi)\right\}\nonumber \\
& &+\Gamma^2\left\{\sin\psi\partial_r+\frac{\cos\psi}{r}(\partial_\psi+i{\bf
A}_\psi)\right\}\nonumber \\
& &+\Gamma^\alpha e^m_a\left(\partial_m+\frac{1}{4}\omega_{m\beta\gamma}
\Gamma^{\beta\gamma}+i{\bf
A}_m\right)\,.
\label{16}
\end{eqnarray}
In this expression Greek indices represent local orthogonal coordinates of $S^2$ and
Latin indices run over internal coordinates. The spin matrices are defined as
$\Gamma^{\alpha\beta}=1/2[\Gamma^\alpha, \Gamma^\beta]$.

We naively expect that, in the large space perspective, the zero modes of $D\!\!\!\!/\,{}_T$ are
the zero-energy modes in the Kaluza-Klein's sense. The crucial condition for the
existence of the zero-energy modes is that the fermion in the large four dimensions
is massless at the center of the core. That is to say, when $\Phi=0$, the four dimensional
mass term should vanish in the present model. Then we first show that the fermion
has massless modes at the axis in four-dimensional effective theory.\cite{11} By the
masslessness in six dimensions, we may consider each member of the fermion in the
doublet of $SU(2)$ to have a positive eigenvalue of $\Gamma_{\#}$ , i.e.,
\begin{eqnarray}
\Psi^1&=&\hat{\chi}\otimes\left(
\begin{array}{c}
1\\0
\end{array}
\right)\,,\nonumber \\
\Psi^2&=&\hat{\lambda}\otimes\left(
\begin{array}{c}
1\\0
\end{array}
\right)\,,
\label{17}
\end{eqnarray}
where $\hat{\lambda}$ and $\hat{\chi}$ are four-component spinors. In the six-dimensional
point of view, the theory is chiral, i.e., $\Gamma_{\#}\Psi=+\Psi$.

Let us consider the following form for the four spinors
\begin{eqnarray}
\hat{\chi}&=&\chi(x)\,e^{-i\phi/2}\,,\nonumber \\
\hat{\lambda}&=&\lambda(x)\,e^{i\phi/2}\,,
\label{18}
\end{eqnarray}
where $\lambda(x)$ and $\chi(x)$ are independent of the internal coordinates, $(\theta, \phi)$.
For these spinors, by inserting Eq.~(\ref{2}) and $\omega_{\phi 45}=-\cos\theta$ into
Eq.~(\ref{12}), the four-dimensional mass term at the axis is written as
\begin{equation}
D_{\phi}\Psi=\frac{i}{2b}\cot\theta\left(
\begin{array}{c}
-(1+\gamma_5)\chi(x)\,e^{-i\phi/2}\\
(1-\gamma_5)\lambda(x)\,e^{i\phi/2}\end{array}\right)\,.
\label{19}
\end{equation}
In this mass term, we may select the large four-dimensional chiralifies of $\lambda$ and $\chi$
as
\begin{equation}
\gamma_5\lambda=\lambda\,,\quad\gamma_5\chi=-\chi\,.
\label{20}
\end{equation}
Then the mass term vanishes. So we find that there exists a massless mode in the
presence of $SU(2)$ monopole con6guration on $S^2$. Furthermore we find that, in the
effective theory, both left- and right-handed massless modes appear.

We now show that the Dirac equation for $\lambda$ and $\chi$ has a solution of bound states
in the core, which are similar to that of Ref.~\cite{8}. The Dirac equation for the base of
$SU(2)$ doublet (\ref{20}) is written as
\begin{eqnarray}
& &~i\sigma^\mu\left(\partial_\mu+\frac{i}{2}A_\mu\right)\chi+i\Phi\lambda=0\,,\nonumber \\
& &-i\tilde{\sigma}^\mu\left(\partial_\mu-\frac{i}{2}A_\mu\right)\lambda+i\Phi^*\chi=0\,,
\label{21}
\end{eqnarray}
where $\sigma^\mu=(1_2, \sigma^i)$ and $\tilde{\sigma}^\mu=(1_2, -\sigma^i)$ $(i=1, 2, 3)$ in
an appropriate representation of $\gamma^\mu$. Note that, the Yukawa couplings in this model
consist of different members of $SU(2)$ doublet. We may factorize the $(z, t)$-dependence of
each two-component ``Weyl'' spinor as
\begin{eqnarray}
\chi&=&\left(\begin{array}{c}
\alpha_1(z+t)\chi_1(r, \psi)\\
\alpha_2(z-t)\chi_2(r, \psi)
\end{array}\right)\,,\nonumber \\
\lambda&=&\left(\begin{array}{c}
\alpha_2(z-t)\lambda_2(r, \psi)\\
\alpha_1(z+t)\lambda_1(r, \psi)
\end{array}\right)\,.
\label{22}
\end{eqnarray}
In this factorization, $\alpha_1$ and $\alpha_2$ are solutions to the equations $(\partial_t-
\partial_z)\alpha_1=0$ and $(\partial_t+\partial_z)\alpha_2=0$, respectively. If this is the
case, the equations of motion are decoupled into two sets $(\chi_1, \lambda_1)$ and $(\chi_2,
\lambda_2)$, so we now consider one of the sets $(\chi_1, \lambda_1)$.

We give an ansatz for their angular dependences:
\begin{equation}
\chi_1=e^{im\psi}\chi(r)\,,\quad\lambda_1=e^{i(m-n+1)\psi}\lambda(r)\,,
\label{23}
\end{equation}
where $m$ is an integer. Then the equations of motion are reduced to
\begin{eqnarray}
& &\left(\partial_r-\frac{m-A_\psi/2}{r}\right)\chi(r)-X\lambda(r)=0\,,\nonumber \\
& &\left(\partial_r+\frac{m-n+1-A_\psi/2}{r}\right)\lambda(r)-X\chi(r)=0\,.
\label{24}
\end{eqnarray}
The asymptotic behaviors of $\lambda$ and $\chi$ at infinity are $\sim e^{\pm r}$, for the bound
states we choose the solution $\sim e^{-r}$. Now we can easily find the number of such zero
modes.
The behaviors of $X$ and $A_\psi$ near $r=0$ are $\sim r^{|n|}$ and $\sim 0$ respectively, then
we find the behavior of $(\chi(r), \lambda(r))$ near the axis as follows,
\begin{equation}
\chi(r)\sim r^m\,,\quad\lambda(r)\sim r^{-m+n-1}\,.
\label{24}
\end{equation}
The regularity condition at $r=0$ is $n-1\ge m \ge 0$. Therefore, the number of zero
modes $(\chi_1, \lambda_1)$ is $n$ in the $n>0$ vortex sector, whereas in the case $n<0$,
there is no regular solution for $(\chi_1, \lambda_1)$. Alternatively, the solutions for
$(\chi_2, \lambda_2)$ exist only in
$n<0$ vortex sector. In both cases, there are $|n|$ normalizable zero modes.

We now discuss the properties of these zero modes.

As the Higgs mechanism, in which the scalar field vacuum expectation value
gives the fermion mass, the fermion in the present model is massive in the region of
``true'' vacuum, in which $SU(2)$ symmetry is restored. This mass term essentially
comes from the compactness of the extra space from the six-dimensional view point
and is of order Planck scale $\sim 1/b$. One set of the zero modes has a freedom of only
one massive fermion, namely, one $SU(2)$ chiral doublet looks like a supermassive
Dirac fermion far from the axis.

In the ordinary four-dimensional superconducting string model of $U(1)\times U(1)$
theory,lo an axial vector coupling is closely connected with the superconductivity of
the vortex. We should mention that the present $SU(2)$ model with Weyl fermion
may have both gauge and gravitational anomalies. Here we do not consider the
cancellation of the anomalies. Similarly, to the $U(1)\times U(1)$ model we expect the
gauge anomaly to be related to the superconductivity. Presumably, it will be
possible that this $SU(2)$ vortex shows the superconductivity by an appropriate
assignment of charges and chiralities of the fermions.

Furthermore, the superconducting string with a bosonic carrier is also possible.
It will be able to construct such a model in the present higher-dimensional vortex
configuration. To this end, we may introduce an extra scalar field which couples
with $SU(2)\times U(1)$ gauge fields. The four-dimensional effective theory resembles
the model in Ref.~\cite{12}, but it may have different properties of detailed structure of
the theory. The contents of this model will be described elsewhere.

To summarize, we have verified the presence of the fermion zero modes
associated with the vortex gauge field on $M_4\times S^2$ which the present authors
recently constructed. Not quite the same as the Jackiw-Rossi type zero modes in
$2+1$ dimension, the topological index is simply calculated as an instanton number.
We have used it and confirmed explicitly that in the $n$-vortex sector there
are $|n|$ normalizable zero modes.

\section*{Acknowledgments}
The authors would like to thank T. Koikawa for reading this manuscript. They
also thank S. Hirenzaki for discussion. This work is supported in part by a Grant-in-Aid for
Encouragement of Young Scientists from the Ministry of Education, Science and Culture (\#
63790150). One of the authors (K. S.) would like to thank the Japan Society for Promotion of
Science for the fellowship. He also thanks Iwanami F\=ujukai for financial aid.


\end{document}